\documentclass[galaxies,article,accept,moreauthors,pdftex]{Definitions/mdpi} 

\usepackage{color}
\usepackage[normalem]{ulem}

\definecolor{darkgreen}{rgb}{0.13, 0.55, 0.13}

\definecolor{brown}{rgb}{0.59, 0.29, 0.0}
\definecolor{ab}{rgb}{0.36, 0.54, 0.66}

\newcommand\aap{Astron. Astrophys.}                
\newcommand\aaps{Astron. Astrophys. Suppl. Ser.}              
\newcommand\aj{Astron. J.}                   
\newcommand\apj{Astrophys. J.}                 
\newcommand\apjl{Astrophys. J. Lett.}                
\newcommand\apjs{Astrophys. J. Suppl.}               
\newcommand\araa{Annu. Rev. Astron. Astrophys.}             
\newcommand\mnras{Mon. Not. R. Astron. Soc.}             
\newcommand\nat{Nature}              
\newcommand\ssr{Space Sci. Rev.}     

\firstpage{1} 
\pubvolume{1}
\issuenum{1}
\articlenumber{0}
\pubyear{2021}
\copyrightyear{2021}
\externaleditor{Elena Moretti and Francesco Longo  
} 
\datereceived{16 July 2021} 
\dateaccepted{1 October 2021 } 
\datepublished{} 
\hreflink{https://doi.org/} 

\Title{Progenitors of Long-Duration Gamma-ray Bursts}

\TitleCitation{Progenitors of Long-Duration Gamma-ray Bursts}

\Author{Arpita Roy 
 \orcidA{}}

\AuthorNames{Arpita Roy}

\AuthorCitation{Roy, A.}

\address[1]{
Cosmology Research Group, Scuola Normale Superiore,
 Piazza dei Cavalieri 7, 56126 Pisa, Italy; arpita.roy@sns.it}

\abstract{We review the current scenario of long-duration Gamma-ray burst (LGRB) progenitors, and in addition, present models of massive stars for a mass range of $10\text{--}150 \, \mathrm{M}_\odot$ with $\Delta \mathrm{M}=10 \, \mathrm{M}_\odot$ and rotation rate $v/v_{\rm{crit}}=0$ to $0.6$ with a velocity resolution $\Delta v/v_{\rm{crit}} =0.1$. We further discuss possible metallicity and rotation rate distribution from our models that might be preferable for the creation of successful LGRB candidates given the observed LGRB rates and their metallicity evolution. In the current understanding, LGRBs are associated with Type-Ic supernovae (SNe). To establish LGRB-SN correlation, we discuss three observational paths: (i) space-time coincidence, (ii) evidence from photometric light curves of LGRB afterglows and SN Type-Ic, (iii) spectroscopic study of both LGRB afterglow and SN. Superluminous SNe are also believed to have the same origin as LGRBs. Therefore, we discuss constraints on the progenitor parameters that can possibly dissociate these two events from a theoretical perspective. We further discuss the scenario of single star versus binary star as a more probable pathway to create LGRBs. Given the limited parameter space in the mass, mass ratio and separation between the two components in a binary, binary channel is less likely to create LGRBs to match the observed LGRB rate. Despite effectively-single massive stars are fewer in number compared to interacting binaries, their chemically homogeneous evolution (CHE) might be the major channel for LGRB production.
}

\begin{document}
\section{Introduction}
\label{intro}

Gamma-ray bursts (GRBs) are intense flashes of high energy electromagnetic radiation of $\sim $ a few 100 keV with a very brief duration of 1 s to a few minutes reaching Earth isotropically from unpredictable directions. GRBs are observationally classified in two groups: short-duration hard ($\sim$1 s, $350$ keV) and long-duration soft ($\sim$a few 10 s of seconds to $\sim$1 min, $\sim $200 keV). These two classes are assumed to have different physical origins. In general, a small amount of matter is required to be accelerated to ultra-relativistic speeds and beamed at a small solid angle to produce GRBs. After their discovery in 1973 \citep{klebesadel1973}, understanding the origin of GRBs has been of utmost importance to comprehend the cosmic evolution of the early universe. Given the association of GRBs with the death of massive stars and them being observable in high redshift, in principle as high as $z \sim 20$ \citep{wijers1998}, GRBs are often considered to probe the star formation histories over cosmic time \citep{bromm2004}.

Massive stars ($\gtrsim 10 \, \mathrm{M}_\odot$) enter the Wolf-Rayet (WR) phase towards the end of or after the main-sequence \citep{roy2020}. Depending on the spectroscopic identification  of heavy elements, such as helium, nitrogen, carbon, oxygen \citep{roy2020, roy2021}, and their excitation states, WR stars are primarily classified in three categories: WN, WC, and WO (see Table 4 of \cite{roy2021}). Theoretically, WN and WC stars are believed to be progenitors of Type Ib and Ic core-collapse supernovae (SNe), respectively, given their pre-SN He- / CO- core masses and the absence of surface H and H/He \citep{woosley2006}. However, there is no direct evidence that suggests single WR stars as Type Ib/Ic SNe progenitors. Given the WR lifetimes of a few 10$^5$ years, one would need to observe $\sim 10^4$ single WR stars to draw a firm connection between them and core-collapse SNe (CCSNe), in a timescale of a few 10s of years. This considerably high number means these are not field stars; rather they are in clusters. Given that more than $70 \%$ of  massive stars are in binaries, the lower mass interacting binaries might be alternative progenitors of SN Type Ib/Ic \citep{crowther2007}.

Energetically, long-duration GRB (LGRB) and CCSNe both fall under the same category. In many long-duration GRBs (LGRB), the beaming-corrected total Gamma-ray energy is estimated to be $\sim 10^{51}$ erg. In addition, the total kinetic energy of the core collapse SN is $\sim 10^{51}$ erg, comparable to that of GRB jets. This naturally makes one consider the possibility of finding the connection between these two extreme phenomena. To confirm this hypothesis, there are three primary approaches available. Firstly, one can find the causal connection---whether both SN and GRB are coincident in space and time. Secondly,  one can study photometry to obtain any overlap in the SN and GRB afterglow light curves. Finally, one can study spectroscopy to establish the SN-GRB connection.

Another class of SNe, Type I superluminous supernovae (henceforth SLSNe), is also believed to have a similar origin as LGRBs. SLSNe are characterized by luminosities $10\text{--}100$~times larger than ``typical” SNe \citep{galyam2012, howell2017, inserra2019}, and their $^{56}$Ni mass of $\sim 20\text{--}30 \, \mathrm{M}_\odot$ is much higher compared to $1 \, \mathrm{M}_\odot$ for ``typical” SNe (\cite{aguilera2020} and references therein). Their spectra show the absence of H and He, same as Type-Ic SN, and they are bare carbon and oxygen cores \citep{galyam2019}. This indicates that SLSNe progenitors have gone through intense mass-loss and/or mixing of chemical elements that made their envelopes depleted of H and He. However, the nature of SLSNe progenitors is yet unknown. Although the most commonly believed theory for their progenitors, based on the observed properties of SLSNe, is the magnetar model---where the newly formed millisecond magnetar, i.e., rapidly rotating, highly magnetized neutron stars (NS), deposits continuous energy to the SN ejecta. LGRBs, on the other hand, are formed in the framework of collapsar model \citep{woosley1993}---where the rapidly rotating stellar core of a massive star collapses into a black hole (BH). Determining the final fate of a massive star that forms a NS or a BH is a complex and poorly understood astrophysical problem. Several recent theoretical models invoke a few diagnostic parameters of the progenitors at the pre-SN or pre-collapse phase that determine the final fates \citep{oconnor2011, sukhbold2014, muller2016, sukhbold2018}. For example, one such diagnostic parameter is the ``so-called” core compactness parameter, $\xi_{\mathrm M}$~\citep{oconnor2011}. We discuss these criteria later in this paper.

This paper aims to review the possible progenitors of LGRBs, find the connection between CCSNe and GRB, and dissociate progenitors of LGRB and SLSN based on several theoretical constraints on the pre-SN cores. Finally, we present a set of models of massive stars for varying mass, rotation rate and metallicity to narrow down a range of values that are favoured by the observed LGRB rate and its metallicity evolution. This paper is organized as follows: in Section \ref{sn-grb}, we discuss the existing observations that associate SNe with LGRBs; in Section \ref{models}, we present the leading models for the progenitors of SLSNe and LGRBs; in Section \ref{single_binary}, we illustrate whether single or binary stars are more suitable candidates for LGRB progenitors; in Section \ref{WR}, we describe the properties of WR stars that are required to form LGRB progenitors, and in Section \ref{disc}, we summarise the salient points of this review article and pose the open questions that will be the focus of research for LGRB astrophysics in the coming years.

\section{SN-GRB Connection}
\label{sn-grb}
\noindent {\bf Space-time coincidence:
} Despite the striking similarity in kinetic energy of SNe and Gamma-ray in GRBs, astronomers did not consider any relation between them \citep{colgate1968, paczynski1986} for decades, only because it was difficult to find out the exact location and thus luminosity of GRBs until the late 1990s. After the confirmation of cosmological length scales and the localization of long-wavelength counterparts, it became more evident that GRBs are associated with young star-forming regions of distant galaxies \citep{costa1997, paradijs1997}, rather than being part of the old galaxies as was previously hypothesized by merger theories (\cite{woosley2006} and references therein). The strong connection of SN-GRB was first established after the discovery of GRB980425 in conjunction with the most unusually bright SN 1998bw \citep{galama1998}, both SN and GRB were coincident in space and time. Another SN 2003dh was also found to be correlated with GRB 030329. Unfortunately, it is difficult to associate each GRB with active star-forming regions in high-redshiftgiven the current instrumental limitations of not resolving $\lesssim 100$ pc at $\sim 100$ Mpc, although the statistical studies reveal a strong correlation between GRBs and blue active star-forming regions of galaxies (\cite{woosley2006} and references therein).

\vspace{5pt}
\noindent {\bf Evidence from photometric light curves:
} 
Observation of GRB 980326 \citep{bloom1998} at redshift unity showed red-emission bump in the optical afterglow \citep{bloom1999a, castro-tirado1999}. This bump in optical was hypothesized to be caused by a consecutive SN followed by the GRB event. The data of red-emission were also consistent with dust re-radiation \citep{waxman2000} from the surrounding material of GRB980326, which again supports the hypothesis of a consecutive SN event. Also, a reanalysis of the GRB970228  afterglow showed the signature of ``bump” rising at a similar time as GRB 980326 \citep{reichart1999, reichart2001, galama2000}. It was difficult to confirm the absolute magnitude of the peak and the type of SN without the spectroscopic redshift of GRB and multi-band photometry of the bump. Future multi-epoch ground- and space- based observations of several GRBs confirmed that the red-emission bump  is indeed associated with SNe \citep{price2003, stanek2005} followed by a GRB event.

The brightness of the GRB 980326 ``bump” matched with typical Type Ic supernova SN 1998bw. Most of the long-soft GRBs are accompanied by Type Ic SNe \citep{woosley2006}. Studies of these SNe show high velocity ejecta causing broad emission lines, and therefore these SNe are classified as ``Type Ic-BL'' \endnote{`BL' stands for broad line. The subclassification of ``Type Ic-BL” is purely based on observations, there is no model to predict its progenitor \citep{woosley2006}.}. However, there are a few exceptions. For example, GRB 060614 is a LGRB with a duration of 102 s, but it has no SN counterpart \citep{fynbo2006, dellavalle2006, galyam2006}. Instead, the re-analysis of its optical afterglow was identified as ``so-called” kilonova emission associated with the compact object merger origin \citep{yang2015, jin2015}. Despite the few exceptions, in most cases, LGRBs are found to be associated with Type Ic-BL SNe. Nonetheless, even though most LGRBs are associated with Type Ic SNe, not all Type Ic SNe are correlated with LGRBs. The reason for different stars to follow different paths is rotation, mass, and metallicity. GRBs are produced by rapidly rotating massive stars that end up with sufficiently rotating pre-SN cores that can create an accretion disk while collapsing into BHs. To retain enough rotation until the pre-SN phase, these stars must not have strong mass loss, and that can easily be achieved at low-metallicity environments. Therefore, metal-poor stars are favoured for the creation of LGRBs. On the other hand, $\sim 66\%$ core-collapse SNe originate from massive stars irrespective of their rotation rates and therefore can happen at both low and high metallicity \citep{woosley2006}.

\vspace{5pt}
\noindent {\bf Spectroscopy:} GRB host galaxies have higher star-formation rate $\sim$a few 10 s of M$_\odot \, \mathrm{yr}^{-1}$, larger than typical field galaxies, as determined by sub-mm observations \citep{berger2003a} and by [Ne III] to [OII] line ratio \citep{bloom2001}. At high redshift ($z$), GRBs track the global star-formation rate \citep{loredo1998, bloom2003, jakobsson2005}. The observations of GRBs on a large scale (high-$z$) confirms predictions of star-formation on small scales. Hence, GRBs, in general, are considered to be good tracers of active star-forming regions. Moreover, the absorption-line spectroscopies give the metallicity estimates of the GRB host galaxies, or in general, the regions through which GRB afterglows are viewed.

Even though there were several GRBs observed whose ``bumps” showed characteristics of Type Ic SN, the solid SN-GRB connection was made after the discovery of the low redshift, $z \sim$0.169, GRB 030329 \citep{greiner2003c} and the accompanied SN 2003dh. Detailed spectroscopy of this GRB afterglow \citep{stanek2003, hjorth2003} showed the deviation from a pure power-law and also a broad SN spectral feature. SN 2003dh spectroscopy was studied in detail as the afterglow faded, and it showed striking similarity with SN 1998bw. Broad spectral lines indicating high velocities $\sim$25,000 km s$^{-1}$ were observed along with the absence of hydrogen, helium and Si II 6355 absorption lines confirming this SN as Type Ic-BL. There were a few other spectroscopic SN-GRB association: for example, at redshift $z= 0.1055$ GRB 031203 \citep{prochaska2004a} accompanied with SN 2003lw \citep{cobb2004}, SN associated with GRB021211 at $z = 1.006$ \citep{della2003}, SN with Swift burst GRB050525a \citep{della2006}.

\section{Models for LGRB and SLSNe}
\label{models}
Several observations and theoretical predictions lead to similar origins for both LGRB and SLSNe, given the similarities in their environments, spectra, and energetics \citep{aguilera2020}. For LGRB, one needs a central engine that can drive a collimated relativistic jet that produces beamed emission \citep{rhoads1997, rhoads1999, frail2001} of energetic $\gamma$- rays. The jet typically has a power of $\sim 10^{50}$~erg~s$^{-1}$ within a narrow opening angle of $0.1$ radian. Therefore, models should provide $\gtrsim 10^{52}$~ergs of energy in a wider angle of $\sim 1$ radian for LGRB to be accompanied with SN like SN 1998bw and SN 2003dh \citep{woosley2007} along with the jet emission. This energy is $\gtrsim 10$ times the ``typical” SN energy. The jet head generally travels at subrelativistic speed with power $\sim 3 \times 10^{48}$ erg s$^{-1}$ inside the star, and it takes 8--25 s to reach the surface with power $\sim 3 \times 10^{50}$ erg s$^{-1}$ \citep {zhang2004}. If the jet is interrupted (there are several interruption scenarios, for example, \cite{zou2021}) or its direction changes in that timespan, then the flow will remain subrelativistic, and therefore will not make a LGRB. Hence, theoretical models suitable for LGRB production need to provide $\gtrsim 10^{50}$ erg s$^{-1}$ of relativistic, beamed power for $\gtrsim 10$~s. Considering these constraints, the most acceptable model for LGRB production is the ``collapsar” model \citep{woosley1993} \endnote{{Contrary to the ``collapsar” model, \cite{dar1999, huang2003} proposed a different origin for GRBs. They argued that when a massive star explodes as SN and the stellar core collapses as a neutron star (NS), the NS acquires high velocity ($\sim 500$ km/s) due to a substantial ``kick” at birth, and as a result, a recoiling ultra-relativistic outflow can be launched in the opposite direction. This outflow can be energetic enough ($\sim 10^{52}$ erg) to produce a long GRB.}}, where a rapidly rotating stellar core collapses to a BH. The suitable progenitor for LGRB is the metal-poor, rapidly rotating either single massive star \citep{yoon2005, woosley2006} that undergoes quasi chemically homogeneous evolution due to rotational mixing, or, a massive star in a closely interacting binary \citep{cantiello2007, detmers2008}.

SLSNe progenitors are also not yet understood completely, there are several leading theories: (i) continuous energy injection to the SN ejecta by the spin-down of the newly-formed central millisecond magnetar \citep{kasen2010, metzger2015, nicholl2017}, (ii) accretion of surrounding ejecta onto the central compact object \citep{moriya2018}, (iii) SN ejecta-circumstellar medium interaction \citep{chatzopoulos2012}, (iv) radioactive decay of large amount of $^{56}$Ni (20--30 M$_\odot$) produced by pair-instability explosion in very massive stars \citep{galyam2019}. Amongst them, the widely accepted theory for SLSNe is the ``magnetar” model (for details, see \cite{aguilera2020} and references therein). In the spirit of the magnetar model, a large number of SLSNe light curves have been analyzed to obtain the distribution of ejecta mass, magnetar spin period and the strength of the magnetic field to reproduce the observables \citep{nicholl2017, blanchard2018, blanchard2019, blanchard2020}. The ejecta masses of SLSNe are estimated to be $3.6 \text{--} 40\, \mathrm{M}_\odot$ that is significantly different from Type Ib/Ic ejecta masses, i.e., strictly $> 10 \, \mathrm{M}_\odot$ \citep{aguilera2020}. Estimated magnetar spin period and magnetic fields are 1 to 8 ms and $0.3$ to $10 \times 10^{14} $G, respectively~\citep{aguilera2020}. Ref. \cite{aguilera2020} found that even the SLSNe progenitors are rapidly rotating, metal-poor massive stars.

Therefore, it is difficult to theoretically predict which stars will explode as SLSNe with NS remnant and which ones will collapse as BH. There are five commonly used parameters to determine a star's explodability criteria. Stellar core compactness is one such parameter, as mentioned in Section \ref{intro}, is defined as,
\begin{linenomath*}
\begin{equation}
\xi_{\mathrm M} = {   {\mathrm{M} / \mathrm{M}_\odot} \over {R (\mathrm{M_{baryon}}=\mathrm{M}) /1000 \, \mathrm{km}}   } \, , 
\label{eq:xi}
\end{equation}
\end{linenomath*}
where $R (\mathrm{M_{baryon}}=\mathrm{M})$ is the radius where the progenitor's core baryon mass $\mathrm{M_{baryon}}=\mathrm{M}$. $\xi_{\mathrm M} $ is the collapse indicator in non-rotating star. {\cite{sukhbold2014} found that $\xi_{\mathrm M}$ is well determined at the Lagrangian mass coordinate of 2.5 M$_\odot$ at the core collapse, where the infall velocity in the core reaches 1000 km/s. Thus $\xi_{\mathrm M} $ is typically denoted as $\xi_{2.5}$.} Stellar cores with  {$\xi_{\mathrm 2.5} \lesssim 3.0\text{--}4.5$} explode as SLSNe {(neutrino winds being the cause of the explosion)} leaving behind NS and higher values {$of \xi_{2.5}$} produce BHs~\citep{aguilera2020, ertl2016, muller2016}. The other four parameters~\citep{aguilera2020, ertl2016} are:
\begin{linenomath*}
\begin{equation}
\mathrm{M}_4 = m(s=4) / \mathrm{M}_\odot \, , 
\label{eq:M4}
\end{equation}
\end{linenomath*}
$m$ is the Lagrangian mass at specific entropy (in the units of $k_{\rm{B}}$) $s=4$;
\begin{linenomath*}
\begin{equation}
\mu_4= \left.{    {dm/\mathrm{M}_\odot} \over {dr/1000 \, \mathrm{km}}    }\right\vert_{s=4}    \, \, ,
\label{eq:mu4}
\end{equation}
\end{linenomath*}
where $dm$ is calculated at $\mathrm{M}_4$, in practice, it is set as $dm=0.3 \, \mathrm{M}_\odot$, and $dr$ is the change in radius between $\mathrm{M}_4$ and $\mathrm{M}_4 + dm$; the dynamo-generated magnetic field strength averaged within the innermost $1.5 \, \mathrm{M}_\odot$,
\vspace{6pt}
\begin{linenomath*}
\begin{equation}
\langle B_{\phi} \rangle = {   {\int_0^{1.5 \, \mathrm{M}_\odot} B_\phi (m) dm}   \over {\int_0^{1.5 \, \mathrm{M}_\odot} dm}   } \, \, ;
\label{eq:b_phi}
\end{equation}
\end{linenomath*}
the mass averaged specific angular momentum within the innermost mass M,
\begin{linenomath*}
\begin{equation}
\bar{j}_{\mathrm{M}} = {    {\int_0^{\mathrm{M}} j_{\mathrm{M}} dm} \over  {\int_0^{\mathrm{M}} dm}    }\, .
\label{eq:jM}
\end{equation}
\end{linenomath*}

Typical values of $\langle B_{\phi} \rangle$ are $\sim 10^{14} \text{--} 10^{15}$ G for NS and an order of magnitude higher for collapsars \citep{aguilera2020}. The average specific angular momentum are $\bar{j}_{1.5 \, \mathrm{M}_\odot} \sim 10^{15}$ (within the innermost 1.5 M$_\odot$) and $\bar{j}_{5 \, \mathrm{M}_\odot} \sim 10^{16}$ cm$^2$ s$^{-1}$ (within the innermost 5 M$_\odot$) for NS and BH progenitors, respectively \citep{aguilera2020}.

\section{Single Stars versus Binary Stars as LGRB Progenitors}
\label{single_binary}
Having discussed the connection between LGRB and Type Ic BL SNe, and the physical models that differentiate the formation of SLSNe and LGRBs, in this section, we discuss scenarios where a single and/or a binary system is a likely progenitor of LGRB. For the LGRB association with Type-Ic SN, the star needs to lose the hydrogen and helium envelopes \endnote{However, in Type-Ic SN, some He may still be present, but just not visible.} and have a large production of $^{56}$Ni. The mass loss of a massive star is dependent on its mass, metallicity and rotation rate. The higher the rotation rate, the larger the mass loss \citep{roy2020, roy2021, choi2016}. With mass loss, the angular momentum transported from the core to the surface eventually gets lost to the interstellar medium. This loss in angular momentum reduces the massive star's rotation rate. Although the star will still have to be sufficiently rotating even after the mass loss to produce the accretion disk around the central BH, which is required for the GRB production. Therefore, the right combination of metallicity and rotation rate needs to be satisfied to create LGRB progenitors. For a detailed discussion of the effect of both rotation rate and metallicity, see Section \ref{WR}.

\vspace{5pt}
\noindent {\bf Single stars:}   Both single and binary channels are possible for LGRB production. For counting as a single star, it might be born ``single” or can be a part of a wide binary with no/minimal interaction between the binary components. In the non-rotating massive stars, heavier elements produced by nucleosynthesis in the core are quasi- chemically homogeneously mixed within the inner convective regions. There are outer layers of lighter elements with gradual decline in atomic masses in the radiative outer zones. In most non-rotating stars, the surface abundance does not evolve much from its initial abundance during the MS \citep{roy2020}. In rotating stars, chemical mixing due to several rotational instabilities (see Section \ref{WR} for details) dredges up the elements from the inner convective core to the surface, crossing the radiative barrier. Therefore, a rapidly rotating massive star can be quasi- chemically homogeneously mixed \citep{roy2021, roy2020, yoon2012, woosley2006, yoon2005} without a clear chemical boundary between the inner convective core and the radiative envelope. The chemically homogeneous evolution (CHE) supplies hydrogen to the core for a longer time producing a larger He-rich core compared to the non-rotating star, leaving little or no hydrogen \citep{levan2016, yoon2005}. Therefore, rotating stars form relatively massive cores from modest initial masses \citep{yoon2005}. Chemically homogeneous massive stars are observed in the Milky Way \citep{levan2016}, even though smaller in number compared to the  Magellanic clouds \citep{martins2009}. The major problem with CHE is that these stars lose their angular momentum while shedding off the outer H- and He- layers via line-driven wind mass loss. This channel helps to create Type Ic SNe at the expense of reduced rotation rate that might make it difficult to produce a rotating BH at the centre. Rapid rotators with $v/v_{\rm {crit}} \gtrsim 0.4$ might satisfy the criteria for both Type-Ic SN and centrally rotating BHs that can produce an accretion disk around it. Thus, we conclude that a sufficiently rotating massive single star can be a legitimate channel for the production of  LGRB progenitor.

\vspace{5pt}
\noindent {\bf Binary stars:} Binary channel is important for both short and long GRBs \citep{levan2016}. Binary stars are abundant both at solar and subsolar metallicities. For example, $\sim 60\text{--}80\%$ MS stars in our Milky Way are in binaries \citep{mason1998, kouwenhoven2005, raghavan2010, sana2012, kobulnicky2014}. Given the bias towards metal-poor environments for the production of GRBs, it is of particular interest to study the Large Magellanic Cloud (LMC) stellar populations. Observations of compact star-clusters NGC 1818, NGC 1805, NGC 1831 and NGC 1868 in the LMC show that almost 55--100\% stars are in binaries \citep{levan2016}. Given the majority of the massive stars residing in binaries, it is worth investigating the physical mechanisms for gaining or retaining angular momentum in binary systems.

Similar to single stars, binary routes also transfer angular momentum to the stellar core during the MS. The star can be spun up via mass transfer from the companion star, or it can also be spun down if the star itself loses mass. It is observed that an older star in the binary gets rejuvenated by the mass transfer from its much younger companion \citep{eldridge2009, stanway2016}. In the context of the binary mass transfer, if the more massive companion loses mass, then it again becomes challenging to produce a sufficiently rotating central BH, similar to the issues faced by single stars. If the mass ratio of the primary and secondary is significantly high, then the massive component (secondary) transfers mass on to the less massive one while on the MS and core He burning phase, and in this scenario, a further little amount of mass transfer from the secondary (after its core He exhaustion) to the CO core of the primary makes the less massive primary companion to spin up and explode energetically to eject the common envelope \citep{podsiadlowski2010, levan2016}.

An alternative channel to obtain the necessary angular momentum is the merger of two companions in a binary. The proposed mechanisms for this channel are either the merger of two He cores or a NS / BH with a He core. The latter channel is thought to be the possible cause of the Christmas-day burst, GRB 101225A \citep{thone2011}. In a merger event, the orbital angular momentum of the two components of the binary is combined in a single merged object. In the case of He core-He core merging, there is a little time lapse between the merging event and SN. In the other scenario of BH-He core merger, BH, in principle, can immediately produce the LGRB via mass accretion from the He-core \citep{levan2016}.

Although there are several channels to produce LGRBs via binary evolution, it is yet not certain if they can produce LGRBs at the necessary rate to match the observed LGRB populations. In each of these binary channels, obtaining the required angular momentum is a major issue, and there is a limited parameter space in initial mass, mass ratio and separation between the two components that can produce the necessary angular momentum (as discussed above) in order to create LGRBs, and therefore, each of these routes contributes to only moderate LGRB-production rates. Hence, even though the majority of the massive stars are in binaries, the binary route of LGRB production is less favoured compared to CHE of single stars to match the observed LGRB production rate, especially at sub-solar metallicity \citep{levan2016}.

\section{Properties of WR Stars Required for LGRB Candidates}
\label{WR}
Following the discussion in Sections \ref{sn-grb} and \ref{models}, it is clear that we need sufficiently rotating massive stars for simultaneous production of LGRBs and Type-Ic SNe. The star should rotate at a speed that helps the stellar core retain enough angular momentum to create an accretion disk when collapsing into a BH, and also produce H- and He- depleted SN. Wind mass loss plays a vital role in determining the end phase angular momentum. Metallicity, being the significant determinant of wind mass loss, along with rotation, decides the final fate of a massive star. Therefore, in this section, we discuss the importance of rotation and metallicity in evolving a massive star into a WN star that eventually might become WC star later in its evolution. The eventual transition to the WC phase is essential because WC stars are believed to be the progenitors of Type-Ic SNe \citep{heger2003, crowther2007} from the theoretical perspective. In this section, we do not directly study the properties of LGRB progenitors and their dependence on mass, rotation rate, and metallicity. We rather follow an indirect approach---we study the mass, rotation rate, and metallicity dependence of massive stars that evolve into WN, and subsequently to WC stars which are SN Type-Ic progenitors. We follow this indirect method because Type-Ic SNe are proxies for LGRB association given their observed correlation.



\vspace{5pt}
\noindent \textbf{Rotation rate:} In this section, we investigate the minimum rotation rate required for an O star to evolve into WN phase. To do that, we study the evolution of surface helium and nitrogen mass fractions for varying mass, metallicity and rotation rate because a certain surface enhancement in He and N makes the transition from O to WN stars. Requirements of He- and N- surface enrichments that determine the WN Late-type (L) phase is given in Table \ref{tab:def_tab_WR}. 

\end{paracol}
\nointerlineskip
\begin{specialtable}[H]
    \tablesize{\small}
    \widetable
\caption{
\label{tab:def_tab_WR}
Taken from \cite{roy2021}. Criteria to classify stars as  O, WNL, WNE, WC, and WO based on the surface enhancement of various elements and core burning status. Surface state and core state refer to conditions at the stellar surface and the centre of the star, respectively; $X_{\rm Q}$ is the mass fraction of element Q. See \cite{roy2021} for detail. 
}
\setlength{\cellWidtha}{\columnwidth/3-2\tabcolsep-0.4in}
\setlength{\cellWidthb}{\columnwidth/3-2\tabcolsep+0.4in}
\setlength{\cellWidthc}{\columnwidth/3-2\tabcolsep-0.0in}
\scalebox{1}[1]{\begin{tabularx}{\columnwidth}{>{\PreserveBackslash\centering}m{\cellWidtha}>{\PreserveBackslash\centering}m{\cellWidthb}>{\PreserveBackslash\centering}m{\cellWidthc}}
\toprule
\textbf{Classification} & \textbf{Surface State} &  \textbf{Central State} \\
\midrule
O stars & $X_{\rm {He}} < 0.4$ &  $X_{\rm{H}}>10^{-4}$ \\
WNL & $X_{\rm{He}}= 0.4 \text{--} 0.9 \qquad X_{\rm C}/X_{\rm N} <10$  & - \\
WNE & $X_{\rm{He}} > 0.9 \qquad X_{\rm C}/X_{\rm N} <10$  & - \\
WC & $0.1<X_{\rm C}<0.6 \qquad X_{\rm O}<0.1 \qquad X_{\rm C}/X_{\rm N} >10$ 
& $X_{\rm{H}}<10^{-4} \qquad X_{\rm{He}}>0.1$\\
WO & $X_{\rm C}>0.1 \qquad X_{\rm O}>0.1 \qquad (X_{\rm C}+X_{\rm O})/X_{\rm{He}} > 1$ & $X_{\rm{H}}<10^{-4} \qquad X_{\rm{He}}<0.1$\\
\bottomrule
\end{tabularx}}

\end{specialtable}
\begin{paracol}{2}
\switchcolumn
Figure \ref{fig:He} shows the surface He mass fraction contours as a function of time and $v/v_{\rm{crit}}$ for a range of mass and metallicity. We notice that all masses show WNL features with surface $^4$He mass fraction $\gtrsim 40\%$ quite early on the MS, $t \lesssim 0.5 \times t_{\rm{MS}}$, if they are moderately or rapidly rotating with $v/v_{\rm{crit}}\gtrsim 0.4$ irrespective of metallicity. This value of $v/v_{\rm{crit}}$ is also supported by the theory of massive star formation \citep{rosen2012}. Therefore, we use $v/v_{\rm{crit}}= 0.4$ as our fiducial case. At solar metallicity, stars show surface enhancement of He when they have lost $\lesssim 50\%$ of their initial masses on the MS irrespective of rotation rate. However, at low metallicities, [Fe/H] $\lesssim -1.0$, we see the surface He enrichment only for $v/v_{\rm{crit}}\gtrsim 0.4$. In these metal-poor moderately or rapidly rotating stars, the rotational mixing of chemical elements dredges up the heavy nucleosynthetic by-products from the core to the surface, even though weak mass loss does not strip off much of their outer layers.   

We show the evolution of surface nitrogen mass fraction in Figure \ref{fig:N}, similar to \mbox{Figure \ref{fig:He}}. The surface N enrichment is consistent with the values inferred for WNL stars. For metal rich stars with [Fe/H] $=0.0$, the surface enrichment is $\sim 17$ times the initial N abundance ($6.73\times 10^{-3}$). This results in surface N mass fraction of 0.011, a factor of 2 less than that is observed in WNL stars in Arches cluster \cite{figer2002}. For metal-poor stars, the surface enrichment is a factor to $28\text{--}30$ compared to their initial abundances of $6.97\times 10^{-4}$, and $7.024\times 10^{-5}$ for [Fe/H] $=-1.0$ and $-2.0$, respectively. 
\begin{figure}[H]
\includegraphics[trim={0.4cm 0cm 0.5cm 0cm}, clip, width=0.75\textwidth]{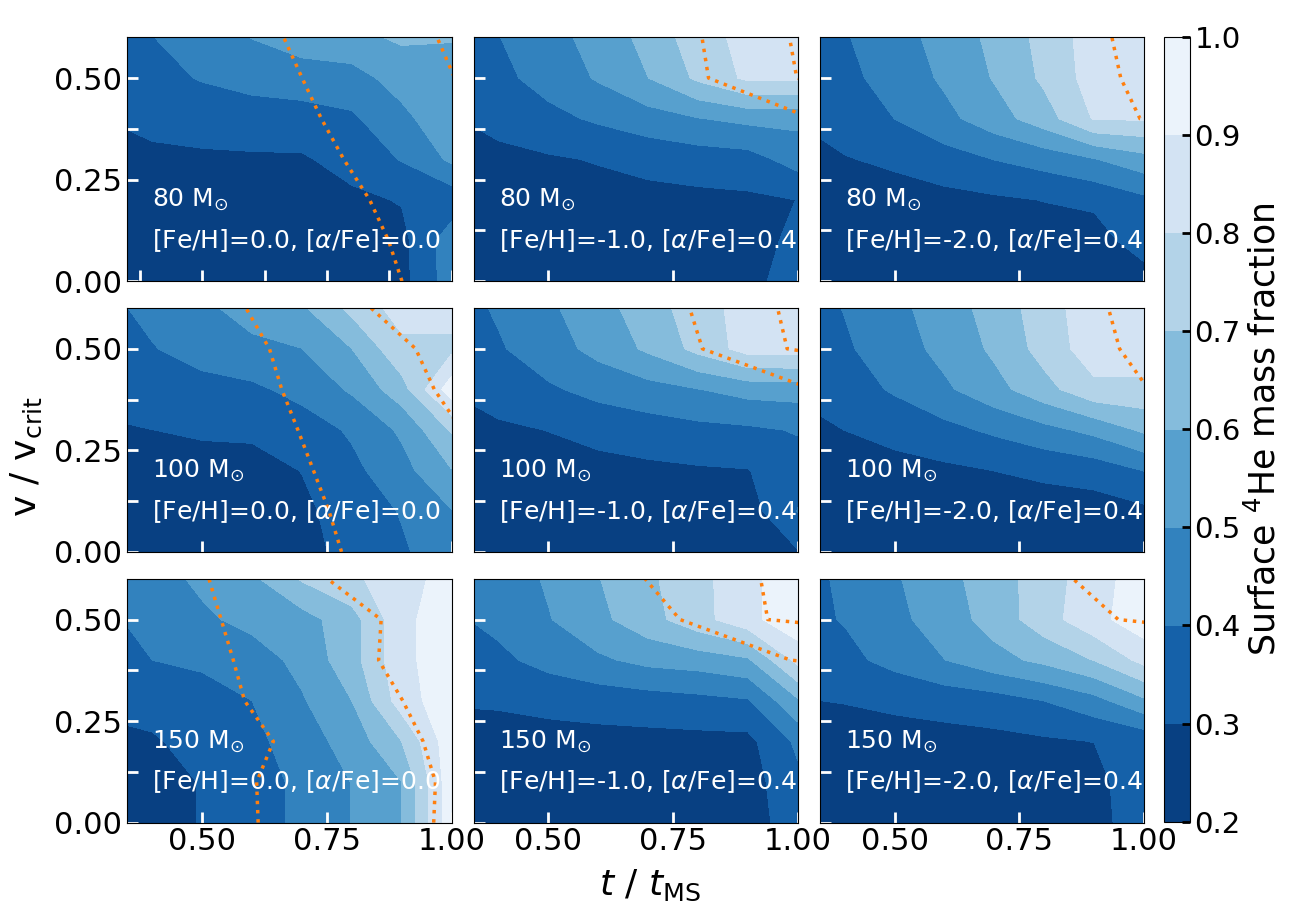}
\caption{$^4$He surface mass fraction, as denoted by the colorbar, as a function of time and rotation rate, $v/v_{\rm{crit}}$ for three metallicities, [Fe/H]  $=0.0$ (leftmost panels), $-1.0$ (middle panels), and $-2.0$ (rightmost panels), and for three masses, $80 \, \mathrm{M}_\odot$ (top panels), $100 \, \mathrm{M}_\odot$ (middle panels), $150 \, \mathrm{M}_\odot$ (bottom panels). We normalize the time by the main-sequence (MS) lifetime, $t_{\rm{MS}}$. The orange dotted lines show the points where 20\% and 50\% of the initial mass are lost. For details, see the discussion of Figure 1 of \cite{roy2020}.
\label{fig:He}}
\end{figure} 

\vspace{-15pt}

\begin{figure}[H]
\includegraphics[trim={0.4cm 0cm 0.5cm 0cm}, clip, width=0.75\textwidth]{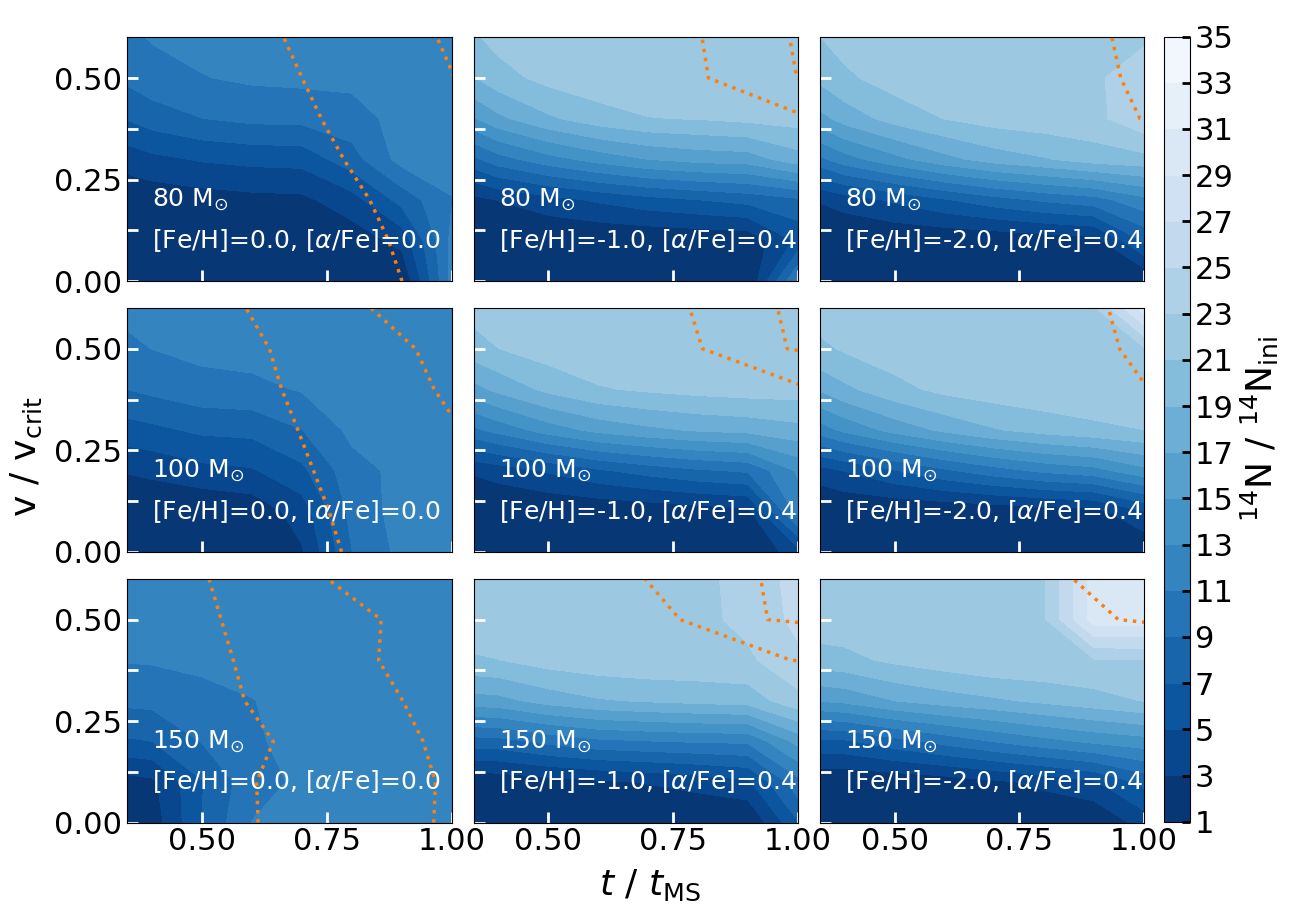}
\caption{Same as Figure \ref{fig:He}, except for the N color contour denoting the $^{14}$N mass-fraction compared to the initial $^{14}$N ($^{14}$N$_{\rm{ini}}$) abundance. $^{14}$N$_{\rm{ini}}$ are $6.73\times 10^{-3}$, $6.97\times 10^{-4}$, and $7.024\times 10^{-5}$, respectively, for [Fe/H] $=0.0$, $-1.0$, $-2.0$, respectively. For details, see the discussion of Figure 2 of \cite{roy2020}.
\label{fig:N}}
\end{figure} 

\vspace{5pt}
\noindent \textbf{Metallicity:} Having discussed the required rotation rate that is favourable for WN and eventual WC production, in this section, we discuss the metallicity range that might be optimum for the production of Type-Ic SN. To study this, we run our models for a mass grid of $10 \, \mathrm{M}_\odot$ to $150\, \mathrm{M}_\odot$ with a mass resolution $\Delta M = 5 \, \mathrm{M}_\odot$ until the end of core $^{12}$C exhaustion ($t_{\rm{C}}$) for our fiducial rotation rate, $v/v_{\rm{crit}}=0.4$, for three metallicities, [Fe/H] $=0$, $-1.0$, $-2.0$, similar to Figure~1 of \cite{roy2021}. We use the 1-D stellar evolution code MESA \citep{paxton2011, paxton2013, paxton2015} Isochrone Stellar Tracks-II (MIST-II \cite{ roy2021, roy2020}, Dotter et~al., 2021, in prep.).
For details of mass-grid and the simulation setup, see~\cite{roy2021, roy2020}.

\begin{figure}[H]
\includegraphics[trim={0.5cm 0cm 0.2cm 0cm}, clip, width=0.75\textwidth]{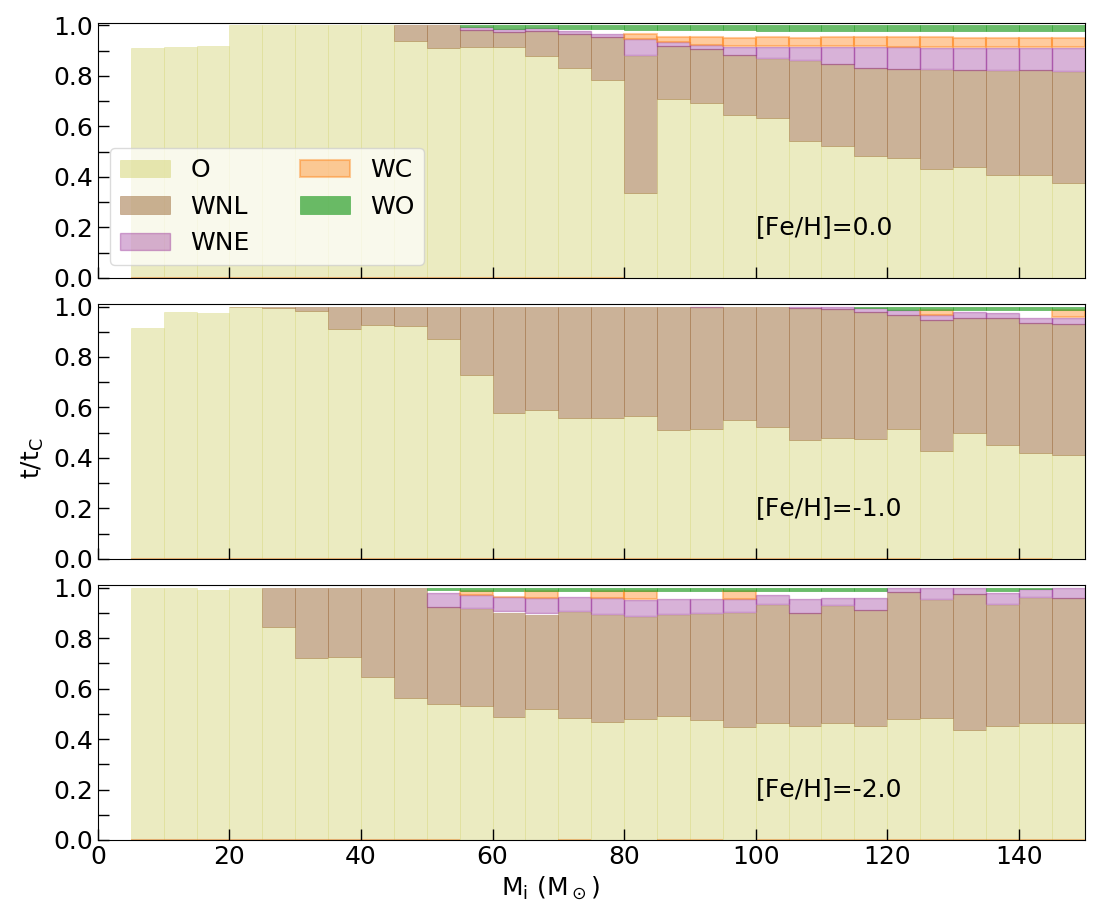}
\caption{The fraction of time each star with a specific intial mass (horizontal-axis) spends in a particular phase classified as O, WNL, WNE, WC, WO, until we stop our simulations at the end of core $^{12}$C exhaustion ($t_{\rm{C}}$), similar to Figure 1 of \cite{roy2021}. These different phases of massive stars are shown by different colors indicated in the figure legend. The definitions for these claasifications of different phases are given in Table \ref{tab:def_tab_WR}. Models shown here are for our fiducial rotation rate, $v/v_{\rm{crit}}=0.4$, and the three panels refer to three metallicities, as indicated.   
\label{fig:WR_class}}
\end{figure} 

\textls[-12]{Definitions of massive star classifications, such as O, WNL, WN Early-type (E), WC, WO, based on surface elemental mass fraction and central burning state, are given in \mbox{Table \ref{tab:def_tab_WR}}. For a detailed discussion of our classifications, see \cite{roy2021}. Here, we briefly} summarize~them:
\begin{itemize}
\item \textbf{O stars.} Stars that are core H burning with mass fraction $X_{\rm{H,\, core}}>10^{-4}$. Nucleosynthetic products do not yet contaminate surfaces of these stars, and therefore, the surface He mass fraction is low, $X_{\rm{He, \, surf}} < 0.4$.
\item \textbf{WNL and WNE stars.} Stars whose surfaces are enriched with He and N due to the mixing of chemical elements driven by various rotational instabilities, and loss of outer envelopes by line-driven wind mass-loss during and after the MS. WNL and WNE stars are defined based on surface abundances of He, N, and C, such as $X_{\rm{He, \, surf}}= 0.4-0.9$ \& $X_{\rm{C, \, surf}}/X_{\rm{N, \, surf}} <10$ and $X_{\rm{He, \, surf}} > 0.9$ \& $ X_{\rm{C, \, surf}}/X_{\rm{N, \, surf}} <10$ for WNL and WNE, respectively.
\item \textbf{WC and WO stars.} Massive stars for which surfaces are contaminated by significant amount of C, later in their evolutions during the late core He burning phase, due to both rotational dredge-up and loss of outer envelopes. These stars might eventually evolve into WO stars given their surfaces have significant amount of O as well. Requirements for surface elemental abundances for the classification of WC and WO are $0.1<X_{\rm{C, \, surf}}<0.6$ \& $X_{\rm{O, \, surf}}<0.1$ \& $X_{\rm{C, \, surf}}/X_{\rm{N, \, surf}} >10$, and $X_{\rm{C, \, surf}}>0.1$ \& $X_{\rm{O, \, surf}}>0.1$ \& $(X_{\rm{C, \, surf}}+X_{\rm{O, \, surf}})/X_{\rm{He, \, surf}} > 1$, respectively. The core H-, He- mass fractions for both these classes are $X_{\rm{H, \, core}}<10^{-4}$, $X_{\rm{He, \, core}}<0.1$.   
\end{itemize}

Having discussed definitions of different classifications of massive stars, we show the fraction of time a star of a given initial mass spends in these individual phases in Figure \ref{fig:WR_class}. We find that stars $< 45 \, \mathrm{M}_\odot$ spend their entire lives as O stars, and beyond this mass, they spend a significant fraction of their lifetimes in the WNL phase for solar metallicity. \textls[-12]{They enter WNL phase for even lower mass stars $\sim 20 \, \mathrm{M}_\odot\text{--}25 \, \mathrm{M}_\odot$ for subsolar metallicities, [Fe/H] $\lesssim -1.0$, independent of metallicity, as can be seen also in Figures \ref{fig:He} and~\ref{fig:N}.} Nonetheless, there is a peculiar metallicity evolution of the stars that show WC and WO features. At solar metallicity, WC and WO features appear for masses $\gtrsim 55 \, \mathrm{M}_\odot$. At subsolar metallicities; however, the minimum mass ($M_{\rm{min}}$) for showing up WC and WO features takes an anomalous turn. It shifts to a larger mass $\gtrsim 115 \, \mathrm{M}_\odot$ for [Fe/H] $=-1.0$, and then again comes down to a lower mass $\sim 50 \, \mathrm{M}_\odot\text{--}55 \, \mathrm{M}_\odot$ for [Fe/H] $=-2.0$. Metal-rich stars undergo strong mass loss, and the mass loss rate decreases with decreasing metallicities. Along with the mass-loss, stars lose angular momentum and therefore slow down, causing less rotational dredge-up at later epochs of their evolutions. Therefore, on the one hand, metal-rich stars, such as solar metallicity stars in our model grid, show heavy metals, such as C and O, on their surfaces when significant mass-loss exposes the metal-rich inner cores~\citep{roy2020} for $M_{\rm{min}} \sim 55 \, \mathrm{M}_\odot$. On the other hand, significantly metal-poor stars, for example [Fe/H] $=-2.0$ in our models, can retain sufficient angular momentum due to weaker mass loss, and therefore might have surfaces enriched with C and O for $M_{\rm{min}} \sim 50 \, \mathrm{M}_\odot$ because of rotational dredge-up, even though they do not expose their metal-rich inner cores. At intermediate metallicity, however, mass loss rates are too weak to expose the metal-rich inner cores but strong enough to lose the angular momentum and therefore to inhibit the chemical mixing driven by rotational dredge-up. This might be the reason why only the most massive stars $\gtrsim 115 \, \mathrm{M}_\odot$ show WC- WO- features at intermediate metallicity [Fe/H] $=-1.0$. This, however, needs to be speculated in detail, and we plan to address this issue in our future paper Roy et~al., 2021, in prep. Note that this result may depend strongly on the adopted mass-loss schemes and their metallicity dependance, and we will explore that in our follow-up paper Roy et~al., 2021, in prep.

Having discussed the metallicity evolution of $M_{\rm{min}}$, we expect to have a larger number of WC and WO stars at solar metallicity, and the number to decrease at [Fe/H] $=-1.0$ and to increase again at [Fe/H] $=-2.0$. Theoretically, WC and WO stars should be progenitors of Type-Ic SN as they lose their H and He envelopes \citep{heger2003, crowther2007}. Therefore, we expect to see a larger number of Type-Ic SNe at solar metallicity decreasing at intermediate subsolar metallicity and increasing again at significantly lower metallicity of [Fe/H]$\sim -2.0$. In addition, there is a one-to-one correlation between LGRB and Type-Ic SNe. We conclude that LGRB numbers starting from solar metallicity might reduce initially with decreasing metallicity and then increase again for significantly metal-poor stars at [Fe/H] $=-2.0$. In metal-poor environments, the LGRB production rate is also favourable because at low metallicities, stellar cores can retain enough angular momentum to form an accretion disk around the collapsing core that creates the central BH. Observations of LGRB also agree with this theoretical prediction that the LGRB rate increases with decreasing metallicities, however, not at excessively low metallicity (e.g., \cite{fruchter2006, graham2013, kruhler2015}). In fact, most observations show that there is a rapid drop-off in the LGRB rate somewhere between solar \citep{kruhler2015, perley2015} and 1/3 solar \citep{graham2015b}, i.e., between the Milky Way (almost solar) and the Small Magellanic Cloud ($\approx$half solar). However, to draw a tight constraint on the metallicity evolution of LGRB rate from both theoretical and observational perspectives, one needs to have observations at metallicities much lower than SMC, and also models with finer metallicity resolutions at these low metallicities. We plan to study this in detail in a follow-up paper Roy et~al., 2021, in prep.

Even though most LGRBs are observed to be associated with SN Type-Ic, there are a few exceptions as discussed in Section \ref{sn-grb}. Therefore, for a conclusive study of LGRB progenitors and their possible connection to SN Type-Ic, we need to investigate the properties of stellar cores and their explodability criteria as given in Equations (\ref{eq:xi})--(\ref{eq:jM}), and also their dependence on mass, rotation rate, and metallicity in detail. We plan to study all these detailed theoretical aspects in a follow-up paper Roy et~al., 2021, in prep.

\section{Summary and Discussion}
\label{disc}
It is almost five decades after the first GRB was discovered in the 1960s, and since then, there are observations of $\sim 1000$ GRBs at different redshifts. These observations have made progress over the years in detecting GRB luminosity and the light curve of the afterglow more accurately, and in advanced spectra aiming to decisively determine the nature of the GRB progenitor and progenitor-circumstellar medium (CSM) interaction. This progenitor-CSM interaction provides hints to the ISM compositions of the host galaxies, and thereby has enriched our understanding of the high-redshift universe. Moreover, the precision to pinpoint the GRB location has also enhanced our knowledge of the host galaxies. A strong correlation between LGRB and Type-Ic SN has broadened our understanding of LGRB progenitors as well.

Even though all these advances have made a clearer picture of GRB production, there are still several open questions yet to be answered:
\begin{itemize}
\item What fraction of O stars really produce GRBs? What does it say about the evolution of their progenitors? 
\item How does the rate of GRB production vary with metallicity and redshift? 
\item {What is the binary fraction that produces LGRBs? How does this fraction vary with metallicity?}
\item {How do we obtain a stronger constraints differentiating LGRB and SLSNe production pathways?}  
\end{itemize}

\funding{This research received no external funding.}

\dataavailability{Data underlying this article will be shared on reasonable request to the author.} 

\acknowledgments{We thank anonymous referees for their insightful comments to enrich the manuscript. AR is grateful to Mark Krumholz and Alexander Heger for their valuable inputs and suggestions. AR gratefully acknowledges the support from Andrea Ferrara's Italian funding scheme ``The quest for the first stars” (Cod. 2017T4ARJ5\_001). AR also gratefully acknowledges support from Mark Krumholz's Australian Research Council's \textit{Discovery Projects} and \textit{Future Fellowship} funding scheme, awards DP190101258 and FT180100375. Parts of this research were conducted by the Australian Research Council Centre of Excellence for All Sky Astrophysics in 3 Dimensions (ASTRO 3D), through project number CE170100013. This research/project was undertaken with the assistance of resources and services from the National Computational Infrastructure (NCI)'s supercomputer Gadi, which is supported by the Australian Government, and the Australian National University's Research School of Astronomy \& Astrophysics's cluster Avatar.}

\conflictsofinterest{The author declares no conflict of interest.}

\end{paracol}
\printendnotes[custom]
\reftitle{References}

\end{document}